# Design of an Encryption-Decryption Module Oriented for Internet Information Security SOC Design

Yixin Liu, Haipeng Zhang, Tao Feng

*School of Electronics & Information, Hangzhou Dianzi University, Hangzhou, China, 310018,* Email: *islotus@163.com*

***Abstract.*** **In order to protect the security of network data, a high speed chip module for encrypting and decrypting of network data packet is designed. The chip module is oriented for internet information security SOC (System on Chip) design. During the design process, AES (Advanced Encryption Standard) and 3DES (Data Encryption Standard) encryption algorithm are adopted to protect the security of network data. The following points are focused: (1) The SOC (System on Chip) design methodology based on IP (Intellectual Property) core is used. AES (Advanced Encryption Standard) and 3DES (Data Encryption Standard) IP (Intellectual Property) cores are embedded in the chip module, peripheral control sub-modules are designed to control the encryption-decryption module, which is capable of shortening the design period of the chip module. (2) The implementation of encryption-decryption with hardware was presented, which improves the safety of data through the encryption-decryption chip and reduce the load of CPU. (3) In our hardware solution, two AES (Advanced Encryption Standard) cores are used to work in parallel, which improves the speed of the encryption module. Moreover, the key length of AES (Advanced Encryption Standard) encryption algorithm is designed with three optional configurations at 128 bits, 256 bits and 192 bits respectively and six optional encryption algorithm modes: CBC (Cipher Block Chaining) mode, ECB (Electronic Code Book) mode, GCM (Galois/Counter Mode) mode, XTS(cipherteXT Stealing) mode, CTR (CounTeR) mode and 3DES respectively, which adds the flexibility to its applications.**

*\*Corresponding address:*
Haipeng Zhang,
 *School of Electronics & Information, Hangzhou Dianzi University,*
*Hangzhou, China,310018*
Email: *islotus@163.com*

## 1. Introduction

Along with the computer information and network technology are widely used in people's life and work, information security is becoming more and more important. Especially in latest years, the development of cloud computing technology identifies the arrival of the age of Internet of Things, which means the security of information and network is much more important [1]. In order to meet the need of network application, a high speed chip for encrypting network data packets is designed. The encryption algorithm is adopt to AES(Advanced Encryption Standard) and 3DES (Data Encryption Standard) encryption algorithm, which are commonly used at present.

With chip's functions becoming more and more complex, explosion of gates on a single chip are caused, and advancement in process technology, requirement for integrating heterogeneous technologies are needed. For the feature size of IC (Integrated Circuit) is nanometer level, therefore it is possible to mount a large and high performance system on one chip and reuse IP (Intellectual Property). IP (Intellectual Property) cores are an important part of the growing EDA (Electronic Design Automation) industry trend towards repeated use of previously designed components [2][3]. We could adopt hierarchical approach and reuse of already designed, optimized and verified design components to meet specification of a complex chip in time and at low cost. About our chip, AES (Advanced Encryption Standard) and 3DES (Data Encryption Standard) encryption algorithm is mature, AES (Advanced Encryption Standard) and 3DES (Data Encryption Standard) IP (Intellectual Property) cores are used in the encryption module, which shortens the development cycle of the chip. The limitation of the current encryption implementation methods is that the throughput is small. One solution to improving the throughput is hardware encryption. The traditional encryption technology is done by running encryption software on the host, which takes up the resources of host exhaustively and operates much slower than hardware encryption. In allusion to this problem, encryption is implemented by hardware. The hardware encryption is independent of the host, and its data storage, operation and so on are realized all through the hardware, so that it has higher speed. The hardware implementation also has better safety [4]. Additional we made two encryption cores working in parallel, which also leads to improved processing speed. In order to meet different need for encrypting different data packet, a variety of encryption modes are designed. Oriented for SOC (System on Chip) and hardware, the encryption module is capable of working at 450MHz and the highest throughput is 11759.616Mbps.

In this paper, we mainly discuss the realization of encryption-decryption module of the internet information security SOC(System On Chip) chip. The paper is organized as follows: in section 2 we introduce at first the two commonly used encryption algorithm AES (Advanced Encryption Standard) and 3DES (Data Encryption Standard). The section 3 is concentrated on the realization of encryption-decryption module. The design architecture is presented, and important sub-modules are discussed in details. In section 4 the experimental results are outlined and compared with those in related references. At last, conclusions are summarized in section 5.

## 2. AES and 3DES algorithm overview

AES(Advanced Encryption Standard) and 3DES (Data Encryption Standard) are commonly used encryption algorithm at present [5].

### 2.1 AES encryption algorithm

AES (Advanced Encryption Standard) encryption algorithm is an iterated block cipher with a variable key length. The key length can be specified to 128, 192 or 256 bits. AES (Advanced Encryption Standard) encryption is composed of key expansion algorithm and encryption (decryption) algorithm, which includes many rounds of iteration and transformation. Procedure may be expressed as follows: key expansion, initialization and iteration rounds. Byte Sub-Transformation, Shift Row transformation, Mix Column transformation and sub-key addition are included in each round except that Mix Column transformation is not included in the last round [6]. In AES(Advanced Encryption Standard) encryption algorithm there are five operation modes. They are CBC (Cipher Block Chaining) mode, ECB (Electronic Code Book) mode, GCM (Galois/Counter Mode) mode, XTS mode and CTR (Counter) mode respectively.

### 2.2  3DES encryption algorithm

DES (Data Encryption Standard) algorithm requires three inputs: lock/key, data and mode. The length of the key is 64 bits. Data is a 64-bit binary. 3DES(Data Encryption Standard) is an encryption algorithm between DES(Data Encryption Standard) and AES(Advanced Encryption Standard) algorithms [7]. That is to say, 3DES(Data Encryption Standard) is a safer DES(Data Encryption Standard) deformation, which is used to design the block encryption algorithm through the combined group method with DES(Data Encryption Standard) as the basic module. Its algorithm architecture can be realized as follows:

Ej() represents encryption process in DES(Data Encryption Standard) algorithm,
Dj() represents decryption process in DES(Data Encryption Standard) algorithm,
K represents the key of DES(Data Encryption Standard) algorithm,
P represents plaintext,
C is for cipher text,
n=1,2,3,… represents the sequence number of the iteration,
Then,
3DES(Data Encryption Standard) encryption process can be expressed as:
$$C = E_{j3}(D_{j2}(E_{j1}(P))); \qquad (1)$$
3DES(Data Encryption Standard) decryption process can be expressed as:
$$P = D_{j1}((E_{j2}(D_{j3}(C))); \qquad (2)$$

## 3. The realization of encryption module in SOC

**Figure1** illustrates the AES(Advanced Encryption Standard) core overview diagram. It consists of 2 Key expansion sub-modules, 2 general AES(Advanced Encryption Standard) encryption sub-modules, dual AES(Advanced Encryption Standard) core control sub-module and miscellaneous AES(Advanced Encryption Standard) modes control sub-module.

The feature of Encryption module:
- Support ECB/GCM/CTR/XTS/CBC/3DES(Data Encryption Standard)
- 450Mhz operation
- Support 128/192/256 bits key for ECB/GCM/CTR/CBC
- Support 256/384/512 bits key for XTS
- Support three 64bits 3DES (Data Encryption Standard)keys and parity bits checking
- 128bits data path for ECB/GCM/CTR/XTS/CBC
- 64bits data path for 3DES(Data Encryption Standard)
- Flow-through design
- Two thread level channels' parallel for ECB/GCM/CTR/XTS
- One engine active for CBC
- Variable performance for 128/192/256 bits Key length

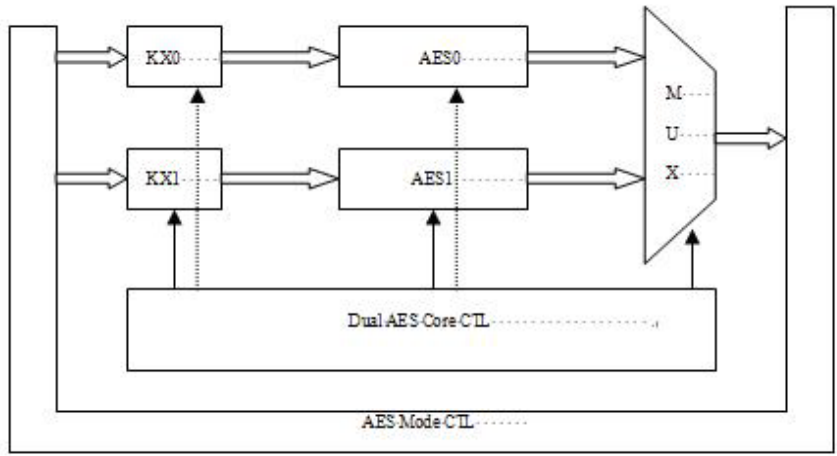

Figure 1. AES Core overview diagram

### 3.1 Key expansion sub-module

The Key expansion sub-module is designed to generate the round key for AES(Advanced Encryption Standard) algorithm. For different key length, the compulsory round for encryption and decryption is different [8]. The key length that most AES(Advanced Encryption Standard) cores support only is the 128 bits and/or 256 bits [9]. Our designed AES(Advanced Encryption Standard) core supports the three input key lengths at 128 bits, 192 bits and 256 bits. Since the data block length is 128 bits according to AES(Advanced Encryption Standard) algorithm, each key is output in 128-bit format each cycle. Both key lengths of 128 bits and 256 bits are multiple of 128, so their expanded keys are easy to be output in 128-bit format. The expansion of 192 bits key can be explained in detail as illustrated in **Figure 2.**

MOD3 CNT is added when in 192 bits key expansion.

In cycle 0, the original 192 bits key is loaded first to the key expansion logic. Then, the first 192 bits round key are calculated by key expansion logic immediately. The round key will be latched by Reg2.9

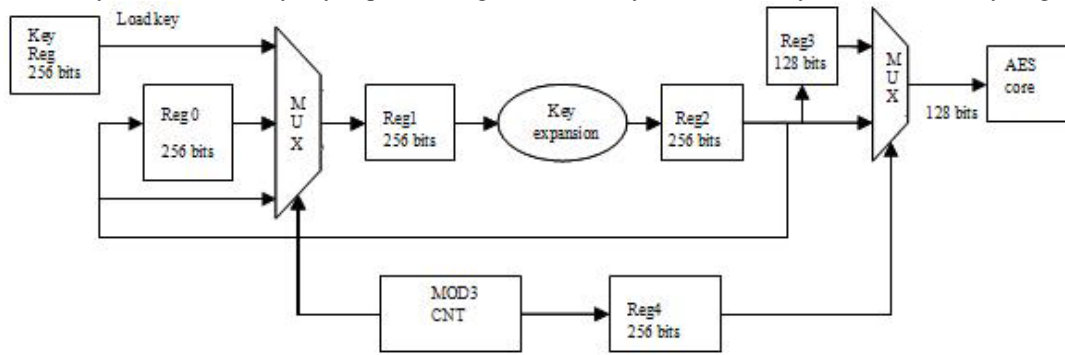

Figure 2. Key expansion sub-module diagram

In cycle 1, the most significant 128 bits of the first round key are output to AES(Advanced Encryption Standard) core through the output stage multiplexer. The least significant 64 bits of the first round key are stored in Reg3. The whole 192 bits of the first round key is input to the key expansion logic for generating the second 192 bits round key. And it is latched by Reg2.

In cycle 2, the least significant 64 bits of the first round key in Reg3 and the most significant 64 bits of the second round key are output to AES(Advanced Encryption Standard) core through the output stage MUX. The remaining 128 bits of the second round key are stored in Reg3. The round key fed into the key expansion logic is halted which is latched by Reg0 in this cycle.

In cycle3, the remaining 128 bits of the second round key are output to AES(Advanced Encryption Standard) core through the output stage MUX. At the same time the second round key are fed to the key expansion logic for generating the third 192 bits round key.

This sub-module is looped to do this operation every three cycles.

### 3.2 AES calculation sub-module

The AES(Advanced Encryption Standard) calculation sub-module is the common sub-module. It is capable of encrypting and decrypting. We input the plaintext and round key and set the encryption mode. This module outputs the internal state result according to the general AES(Advanced Encryption Standard) algorithm.

### 3.3 Dual AES core control sub-module

To improve the performance of the encryption module, two AES(Advanced Encryption Standard) cores are adopted for parallel operation. This sub-module is used to control the multiplexer to output one of the two results properly and control the round number according to the key length, decide whether to do rotation operation when in 256 bit key mode and disable one channel when in CBC mode.

It takes one AES(Advanced Encryption Standard) core 14 cycles to finish a complete AES(Advanced Encryption Standard) operation (for example the key length is 256 bits). The Dual AES(Advanced Encryption Standard) core control logic feeds the first block data to AES0, then 7 cycles later, the second data block will be feed to AES1 except for the CBC mode. Since stateful algorithm is adopted in the CBC mode, only AES0 is enabled in the whole 14 cycles. The second 7 cycles later, the processed result of first block data is output by AES0. At the same time, AES0 is idle and another block data could be fed to AES0. The third 7 cycles later, the input data is processed by AES1. The multiplexer is used to choose the path to output data. This process does not stop until the data stream is end.

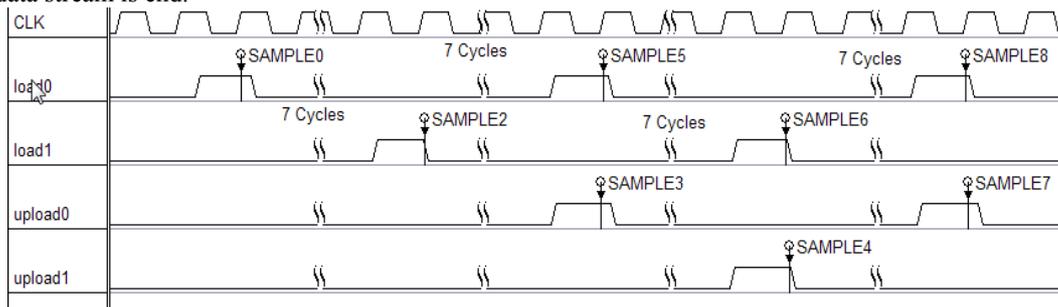

**Figure 3.** Dual AES core Control

### 3.4 AES mode control sub-module

AES(Advanced Encryption Standard) mode control sub-module is a key component of the core. It is used to control the data path for miscellaneous AES(Advanced Encryption Standard) modes. An internal State-Machine is designed to control the whole operation, from loading keys, fetching data to outputting result.

The complete AES(Advanced Encryption Standard) core contains many encryption modes. The AES mode module combines all the modes into a unique unit and fully utilized circuit is shown in **Figure 4**. Many MUX and XOR are applied for specified algorithm.

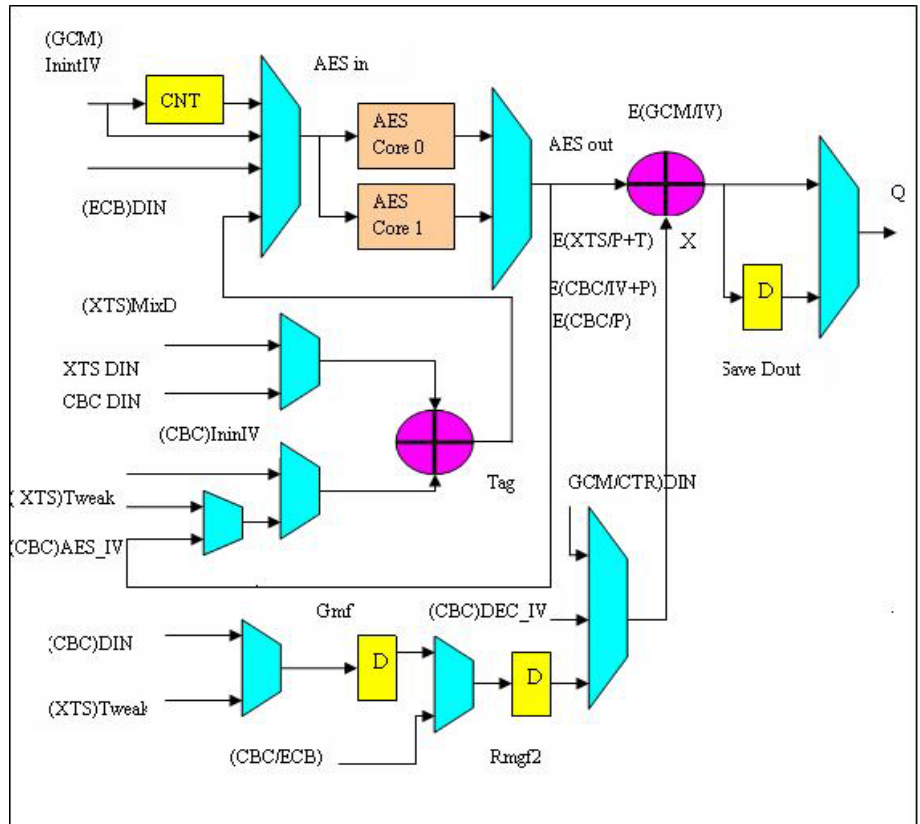

**Figure 4.** Dual AES core Control

## 3.5 Signal description

The complete AES(Advanced Encryption Standard) IP core supports six modes of AES (Advanced Encryption Standard)and 3DES(Data Encryption Standard) encryption algorithms. They are GCM/CBC/CTR/ECB/XTS for AES(Advanced Encryption Standard) and 3DES(Data Encryption Standard). The AES(Advanced Encryption Standard) and 3DES(Data Encryption Standard) are completely different from each other in protecting the original data. Thus to achieve the goal to contain six modes in one IP core is to combine two different encryption cores, which is illustrated in **Figure5** as follows.

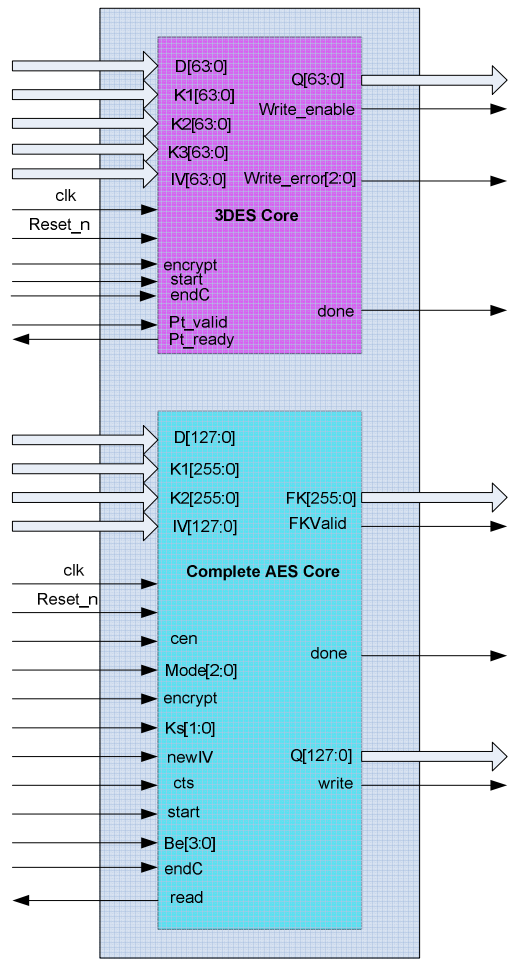

Figure 5. Complete AES core interface diagram

**Table 1** and **Table 2** list out the description of I/O ports for AES(Advanced Encryption Standard) and 3DES(Data Encryption Standard).

Table 1. AES core Description

| Signal Name | I/O Type | Description |
| --- | --- | --- |
| Clk | Input | Core clock signal |
| Reset_n | Input | Core reset signal(Low active) |
| Icg_disable | Input | Global signal to disable internal clock Gating cell inside encryption core |
| Cen | Input | Enable input data, when it's deserted, core will ignore the input data stream |
| Mode[2:0] | Input | Encryption mode:<br>3'b000    GCM<br>3'b001    CBC |
| | | 3'b010   CTR <br> 3'b011   ECB <br> 3'b100   XTS |
| Encrypt | Input | Indicate the current operation is encryption, otherwise the operation is decryption. |
| Ks[1:0] | Input | The operation key size: <br> 2'b00   128bits <br> 2'b01   192bits <br> 2'b10   256bits |
| NewIV | Input | (XTS mode only) <br> Marks the last block of the data sub-module if followed immediately by the first block of the next data sub-module with different IV |
| Cts | Input | (XTS mode only) <br> Marks the last full 128-bit block of the data sub-module in case that the next block of this data sub-module is less than 128 bits |
| Start | Input | Start of input data stream |
| Read | Output | Handshake signal <br> Indicate the core read the data from source |
| D[127:0] | Input | Input data, plaintext for AES algorithm |
| K1[255:0] | Input | 128/192/256 bit AES key <br> (128 bits Key use k1[255:128], <br>    192 bit key use k1[255:64]) |
| K2[255:0] | Input | (XTS mode only) <br> Tweak key(Key2) <br> (128 bit Key use k2[255:128], <br>    192 bit key use k2[255:64]) |
| Iv[127:0] | Input | In GCM/CTR mode: initial counter value. (software will append the left bit to align 128bit boundary) <br> In CBC mode: initial value <br> In XTS mode: 128 bit Tweak for calculation |
| Be[3:0] | Input | (GCM and XTS modes only) <br> Byte length of the last data block in bytes minus 1 <br> 0 – corresponds to 1 byte <br> 1 – corresponds to 2 bytes <br> 15 – corresponds to 16 bytes |
| EndC | Input | (GCM/CTR mode only) <br> Mark the last DATA block |
| Q[127:0] | Output | The result DATA of cipher text |
| Write | Output | Write enable signal for result cipher text |
| FK[255:0] | Output | Final round Key(used for real time verification if possible) <br> (128 bit Key use FK[255:128], <br>    192 bit key use FK[255:64]) |
| FKvalid | Output | Enable the final round key |
| Done | Output | HIGH when data processing is completed |

    The 3DES(Data Encryption Standard) encryption algorithm is completely different from AES(Advanced Encryption Standard). An individual interface for this function block is set in internally so called "complete encryption core".

Table 2. 3DES core Description

| Signal Name | I/O Type | Description |
|---|---|---|
| Clk | Input | Core clock signal |
| Reset_n | Input | Core reset signal(Low active) |
| D[63:0] | Input | Input DATA, plaintext for 3DES algorithm |
| K1[63:0] | Input | 64bits 3DES Key1 |
| K2[63:0] | Input | 64bits 3DES Key2 |
| K3[63:0] | Input | 64bits 3DES Key3 |
| IV[63:0] | Input | 64bits Initial vector number |
| Encrypt | Input | HIGH when core is encrypting. Otherwise the core is decrypting. |
| Start | Input | HIGH when the core starts to do 3DES algorithm. |
| EndC | Input | HIGH when the last DATA block is coming. |
| Pt_valid | Input | HIGH when the plaintext is valid |
| Pt_ready | Output | HIGH when the core begin to get in the plaintext |
| Q[63:0] | Output | The result DATA of cipher text |
| Write_enable | Output | Write enable signal for result cipher text |
| Done | Output | HIGH when data processing is completed |
| Write_error[2:0] | Output | Indicate the parity error for three input keys |

## 4. Experiment results and discussions

Table 3 lists part of the performance indices. Table 4 lists the comparison of our results with other similar published researches. As can be seen in tab 3, the performance of 3DES(Data Encryption Standard) is worse than that of AES(Advanced Encryption Standard). Since AES-CBC is single path, its performance is a little worse than those of other modes, which also proves that two modules working in parallel is capable of improving the performance. We not only use two modules to work in parallel, but also add some software algorithm in key expansion. Seen from tab 4, we realize a better balance among the use of resource, frequency and throughput.

Table 3. Performance table

| algorithm | Key size(bits) | Performance (Mbps) |
|---|---|---|
| 3DES | 192 | 3297.28 |
| AES-CBC | 128 | 6949.888 |
| AES-CBC | 192 | 5886.976 |
| AES-CBC | 256 | 5168.128 |
| AES-CTR | 128 | 11759.616 |
| AES-CTR | 192 | 9803.776 |
| AES-CTR | 256 | 8406.016 |
| AES-ECB | 128 | 12566.528 |
| AES-ECB | 192 | 10478.592 |
| AES-ECB | 256 | 9373.696 |
| AES-GCM | 128 | 11759.616 |
| AES-GCM | 192 | 9803.776 |
| AES-GCM | 256 | 8406.016 |

Table 4. Performance comparison table

| Designer | Frequency (MHz) | Throughput (Mbps) | Gate count (k) | Technology (μm) |
|---|---|---|---|---|
| reference[10] | 125 | 1454 | 173 | 0.18 |
| reference[11] | 131 | 311 | 5398 | 0.11 |
| reference[12] | 104 | 1330 | 15073 | 0.18 |
| reference[13] | 780 | 9076 | 167 | 0.18 |
| reference[14] | 250 | 2900 | 63.4 | 0.25 |
| this article | 450 | 11759.616 | 200 | 0.04 |

## 5. Conclusions

An encryption module consist of 3DES(Data Encryption Standard) and AES(Advanced Encryption Standard) algorithm oriented for internet information security SOC(System On Chip) design is designed, which supports 128 bits, 192 bits, 256 bits key at the same time. Two AES(Advanced Encryption Standard) Cores are utilized to work in parallel, which is proved to improve the speed of the encryption module obviously. The performance of encryption module with different operating modes is tested with VCS simulator. The frequency of encryption module is capable of operating up to 450 MHz with gate count of 200k. By comparing with those results in several similar published papers, our results appear the highest throughput at a middle frequency and our design occupies fewer resources. A better balance among resources occupation, frequency and throughput is abtained in our designed SOC(System On Chip) module. For the CBC mode is stateful, the 2 AES(Advanced Encryption Standard) cores can't work in parallel. As a result, the throughput of CBC mode is lower.

In the future, we will try to explore a method that all the encryption modes are working at high speed include CBC mode.

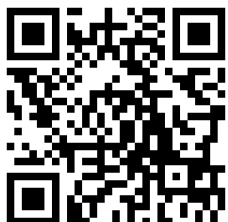

Free download and more information for this paper